\documentclass{ws-ijmpa}

\begin{document}

%%%%%%%%%%%%%%%%%%%%%%%%%%% To switch off trimmarks %%%%%%%%%%%%%%%%%%%
%

\def\nocropmarks{\vskip5pt\phantom{cropmarks}}

%\let\trimmarks\nocropmarks      %%% Pls. remove the comment sign (%) to
%%switch off the trimmarks

%
%%%%%%%%%%%%%%%%%%%%%%%%%%%%%%%%%%%%%%%%%%%%%%%%%%%%%%

\markboth{Detmold, Melnitchouk \& Thomas}
{Connecting Structure Functions on the Lattice with Phenomenology}

%%%%%%%%%%%%%%%% Publisher's Area please ignore %%%%%%%%%%%%%%%
%
\catchline{}{}{}
%
%%%%%%%%%%%%%%%%%%%%%%%%%%%%%%%%%%%%%%%%%%%%%%%%%

\setcounter{page}{1}

\title{CONNECTING STRUCTURE FUNCTIONS ON THE LATTICE WITH PHENOMENOLOGY}

\author{W. DETMOLD$^1$, W. MELNITCHOUK$^2$, A. W. THOMAS$^1$}
\address{$^1$
	Special Research Centre for the Subatomic Structure of Matter,
	and Department of Physics and Mathematical Physics,
	Adelaide University 5005, Australia	\\
	$^2$
	Jefferson Lab, 12000 Jefferson Avenue,
	Newport News, Virginia 23606, USA}

\maketitle

%\pub{Received (received date)}{Revised (revised date)}

\begin{abstract}
We examine the extraction of moments of parton distributions from
lattice data, focusing in particular on the chiral extrapolation as
a function of the quark mass.
Inclusion of the correct chiral behavior of the spin-averaged isovector
distribution resolves a long-standing discrepancy between the lattice
moments and experiment.
We extract the $x$-dependence of the valence $u-d$ distribution from
the lowest few lattice moments, and discuss the implications for the
quark mass dependence of meson masses lying on the $\rho$ Regge
trajectory.
The role of chiral symmetry in spin-dependent distributions, and in
particular the lattice axial vector charge, $g_A$, is also highlighted.
\end{abstract}

%%%%%%%%%%%%%%%%%%%%%%%%%%%%%%%%%%%%%%%%%%%%%%%%%%%%%%%%%%%%%%%%%%%%%%%%%
\section{Introduction}

Parton distribution functions (PDFs) parameterize fundamental
information on the nonperturbative structure of the nucleon, and the
workings of QCD at low energy.
Over the past two decades considerable experience has been gained with
studies of PDFs within low energy models of the nucleon.
Ultimately, however, one would like a more rigorous connection of PDFs
with QCD, and presently this can only be provided through the lattice
formulation of QCD.

Because PDFs are defined as light cone correlation functions, it is not
possible to calculate them directly on the lattice in Euclidean space.
Instead, one calculates matrix elements of local twist-two operators,
which are related through the operator product expansion to moments of
the PDFs.
For the spin-averaged quark distributions,
$q(x) = q^\uparrow(x) + q^\downarrow(x)$,
the moments are defined as:
\begin{eqnarray}
\label{moments}
\langle x^n \rangle_q
&=& \int_0^1 dx\ x^n\
\left( q(x) + (-1)^{n+1} \bar q(x) \right)\ ,
\end{eqnarray}
while moments of the helicity distributions,
$\Delta q(x) = q^\uparrow(x) - q^\downarrow(x)$, are given by:
\begin{eqnarray}
\label{helmoments}
\langle \Delta x^n \rangle_q
&=& \int_0^1 dx\ x^n\
\left( \Delta q(x) + (-1)^n \Delta \bar q(x) \right)\ .
\end{eqnarray}
A number of lattice calculations of PDF moments have been performed over
the last decade, initially in the quenched approximation, and more
recently with dynamical quarks.
The results indicate that at the relatively large quark masses at which 
the calculations were made (between $m_q \approx 30$ and 190~MeV), the
unquenched results are indistinguishable from the quenched within the
current errors.

Despite the impressive progress of the lattice calculations, for many
years the moments $\langle x^n \rangle_q$ have yielded results which
were typically $\sim 50\%$ larger than the experimental values, when
linearly extrapolated to the physical quark masses.
This discrepancy was recently resolved with the observation\cite{DMNRT}
that a linear extrapolation in quark mass omits crucial physics
associated with the nucleon's
% long range structure of the nucleon in the form of
pion cloud, and that inclusion of the nonanalytic dependence on the
quark mass is essential if one is to reconcile the lattice data with
experiment.

In this paper we report recent progress made in connecting moments
of parton distributions calculated on the lattice with experiment,
focusing on the extraction of the $u-d$ moments (which are insensitive
to the poorly known disconnected contributions\cite{DMNRT}).
We discuss the extraction of the $x$ dependence from moments, and the
implications for masses of mesons lying on the $\rho$ Regge trajectory,
and outline the role of chiral symmetry in helicity distributions and
the axial vector charge, $g_A$.

%%%%%%%%%%%%%%%%%%%%%%%%%%%%%%%%%%%%%%%%%%%%%%%%%%%%%%%%%%%%%%%%%%%%%%%%%
\section{Chiral Extrapolation of Lattice Moments}

The spontaneous breaking of the chiral SU(2)$_L\times$SU(2)$_R$ symmetry
of QCD generates the nearly massless Goldstone bosons (pions), whose
importance in hadron structure is well documented.
At small pion masses, PDF moments can be systematically expanded in a
series in $m_\pi$, with the expansion coefficients generally free
parameters.
One of the unique consequences of pion loops, however, is the appearance
of terms nonanalytic in the quark mass, $m_q \propto m_\pi^2$, which
arise from the infrared behavior of the chiral loops, and are therefore
model independent.

The leading order (in $m_\pi$) nonanalytic term in the expansion of the
moments of PDFs was shown by Thomas et al.\cite{TMS} to have the generic
behavior $m_\pi^2 \log m_\pi$.
On the other hand, in the heavy quark limit, in which the valence quark
distributions become $\delta$-functions centered at $x=1/3$, the moments
$\langle x^n\rangle_{u-d}$ approach $1/3^n$.
An extrapolation formula which explicitly satisfies both the heavy
quark and chiral limits can be written:\cite{DMT}
\begin{equation}
\label{xtrap}
\langle x^n \rangle_{u-d}\
=\ a_n \left( 1 + c_{\rm LNA}m_\pi^2\log\frac{m_\pi^2}{m_\pi^2+\mu^2}
       \right)\
+\ b_n \frac{m_\pi^2}{m_\pi^2+\lambda_n^2}\ ,
\end{equation}
in which the coefficient of the leading nonanalytic (LNA) term,
$c_{\rm LNA} = -(1 + 3 g_A^2)/(4\pi f_\pi)^2$, is calculated from chiral
perturbation theory.\cite{SAV,JI}
The mass $\mu$ reflects the scale at which the Compton wavelength of the
pion becomes comparable to the size of the hadron (without its pion
cloud).
Previous fits\cite{DMNRT} to the lattice moments of $u-d$ suggest a
value $\mu \approx 550$~MeV, which is consistent with values
found\cite{OBS} in analyses of nucleon static properties.
The $b_n$ term in Eq.~(\ref{xtrap}),
$b_n = 1/3^n - a_n \left( 1 - \mu^2 c_{\rm LNA} \right)$,
is included in order to provide a linear dependence on $m_\pi^2$, and
the mass scale $\lambda_n$ is set to be 5~GeV for all $n$.

In Fig.~1 we show the best $\chi^2$ fit to the lattice
data\cite{QCDSF,MIT} for the $n=1$ moment of $u-d$ as a function of
$m_\pi^2$.
Clearly, an extrapolation based on Eq.~(\ref{xtrap}) provides a much
better fit to the lattice data and experiment than a linear fit.
Similar results are found\cite{DMNRT} for the $n=2$ and $n=3$ moments.
Unfortunately, because all of the lattice data are in a region where
the moments show little variation with $m_\pi^2$, it is not possible to
determine $\mu$ from the current data, and within the errors both the
lattice data and the experimental values can be fitted with $\mu$
ranging from $\sim 400$ to 700~MeV.
Data at smaller quark masses are therefore crucial to constrain this   
parameter and guide an accurate extrapolation.

\begin{figure}[htbp] %ORIGINAL SIZE: width=1.4TRUEIN;height=1.5TRUEIN
\begin{center}
\epsfig{figure=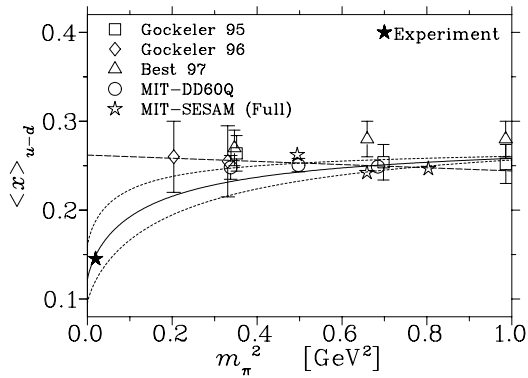,height=5.5cm}
\caption{The $n=1$ moment of the $u-d$ distribution, with a best fit
	to data\protect\cite{QCDSF,MIT} using Eq.(\ref{xtrap}) with
	$\mu=550$~MeV (solid), and $\mu=400$ and 700~MeV (upper and
	lower dotted), compared with a linear extrapolation (dashed).}
\end{center}
\end{figure}

%%%%%%%%%%%%%%%%%%%%%%%%%%%%%%%%%%%%%%%%%%%%%%%%%%%%%%%%%%%%%%%%%%%%%%%%%
\section{Quark Distributions from Lattice Moments}

Having established the appropriate way to extrapolate lattice data on
PDF moments, one can now ask much information can the existing data
provide on the $x$ dependence of the PDFs.
The reconstruction of the complete $x$ dependence in principle requires
infinitely many moments, however, on the lattice, because of operator
mixing for operators with spin $\geq 5$, all calculations have so far
been restricted to $n \leq 3$.
Nevertheless, as shown by Detmold et al.,\cite{DMT} considerable
information on the shape of the valence $u_v-d_v$ distribution can
already be inferred from just the lowest four moments.

The most efficient way to reconstruct the PDF from a limited number of
moments is to determine the parameters of the PDF parameterization by
directly fitting to the moments.
For the isovector valence distribution accurate PDFs can be
reconstructed from the lowest four moments, using the standard
parameterization,
\begin{equation}
\label{param}
x q_v(x)
\equiv x (q(x) - \bar q(x))
= \alpha x^\beta (1-x)^\delta (1 + \epsilon\sqrt{x} + \gamma x)\ ,
\end{equation}
where $\beta$ can be related through Regge theory to the intercept of
the $\rho$ Regge trajectory, and $\delta$ is given by perturbative QCD
counting rules.

Note that some care must be taken when attempting to extract information
on valence quark distributions from both the even and odd moments.
While the even $n$ moments of $u-d$ correspond to the valence
distribution, $u_v-d_v$, the odd $n$ moments correspond to the
combination $u_v-d_v+2 \bar u-2 \bar d$.
The difference between these represents the famous violation of the
Gottfried sum rule.
Given sufficiently many moments of $u-d$, one can reconstruct both the
valence $u_v-d_v$ and $\bar u-\bar d$ distributions by fitting the even
and odd moments separately.
However, at present there exists only a single data point for the
even moments, $n=2$ (the $n=0$ point corresponds to normalization),
which makes it difficult to obtain accurate information.
To minimize the error associated with reconstruction of the
$\bar u-\bar d$ distribution, we subtract the values of the
phenomenological moments of $\bar d-\bar u$ from the calculated odd
moments.
The correction to the $n=1$ moment is $< 10\%$, while for $n=3$ it is a
fraction of a percent.

\begin{figure}[htbp]
\begin{center}
\epsfig{figure=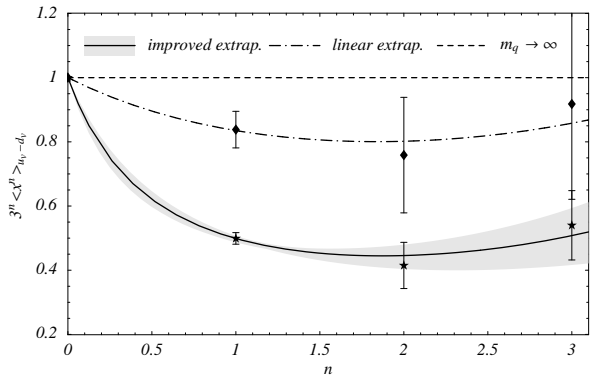,height=5cm}
\caption{Moments of the valence $u_v-d_v$ distribution
	(scaled by $3^n$) at the physical quark mass, extracted from
	fits to lattice data using a linear extrapolation (diamonds)
	and Eq.~(\protect\ref{xtrap}) (stars).}
\end{center}
\end{figure}

The resulting fits to the moments of the valence $u_v-d_v$ distribution
are displayed in Fig.~2 for both the linear and improved extrapolation,
Eq.~(\ref{xtrap}), with the shaded region around the latter corresponding
to a 1~$\sigma$ variation of the fit parameters from their optimal values.
For clarity we plot $3^n$ times the moments, so that the horizontal line 
at unity represents the heavy quark limit.

The corresponding distributions $x (u_v-d_v)$ are displayed in Fig.~3.
Once again, the lightly shaded region represents a 1~$\sigma$ deviation  
from the central values, while the darker band illustrates the spread
between global PDF fits.
A comparison of the distribution reconstructed using the improved chiral
extrapolation with the phenomenological distributions shows reasonably
good agreement.
On the other hand, the linear extrapolation gives a distribution (scaled
by a factor $1/2$ in the figure) which has a significantly higher peak,
centered at $x \sim 1/3$, reminiscent of a heavy, constituent quark--like
distribution.

\begin{figure}[htbp] 
\begin{center}
\epsfig{figure=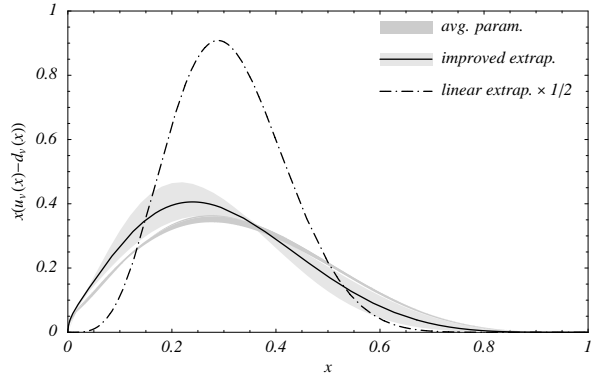,height=5cm}
\caption{Physical valence $x(u_v-d_v)$ distribution, extracted using
	Eq.~(\protect\ref{xtrap}) (solid), and a linear extrapolation,
	scaled by a factor 1/2 (dot--dashed).}
\end{center}
\end{figure}

%%%%%%%%%%%%%%%%%%%%%%%%%%%%%%%%%%%%%%%%%%%%%%%%%%%%%%%%%%%%%%%%%%%%%%%%%
\section{Regge Behavior and the $\rho$ Regge Trajectory}

According to Regge theory, the exponent $\beta$ which governs the
small-$x$ behavior of the distribution $u_v-d_v$ is related to the
intercept ($\approx 1/2$) of the isovector, $C$-odd, $\rho$ Regge
trajectory, and indeed the best fit\cite{DMT} to parameterizations of
global data gives $\beta \approx 0.48$.
The dependence of $\beta$ on the quark mass obtained from the fits to
moments in Sec.~3 therefore allows one to predict the $m_q$ dependence
of the Regge intercept.

In addition to the intercept, one also needs to determine the slope of
the trajectory as a function of $m_q$.
In the infinite mass limit, orbital excitations of mesons become
energetically degenerate with the $L=0$ state.
Within Regge theory, this is possible only if
$\beta \to \infty$ as $m_q \to \infty$, which is consistent with the
valence distribution approaching a $\delta$-function.
One expects, therefore, that the slope should increase as $m_q$ increases
from its physical value.
Using the lattice data for the $\rho$ meson and the values of $\beta$
generated from the fits in Fig.~2, one can then make predictions for the
behavior of the masses of the orbital excitations as a function of $m_q$.

\begin{figure}[htbp]
\begin{center}
\epsfig{figure=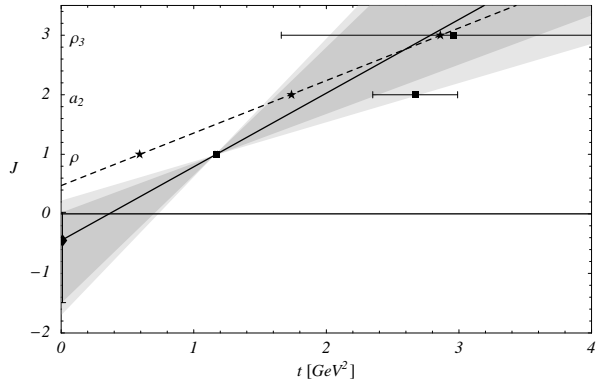,height=5cm}
\caption{Regge plot of the spin, $J$, versus $t=({\rm mass})^2$ of
	mesons on the $\rho$ trajectory, at the physical pion mass
	(dashed) and at $m_{\pi}=0.785$~GeV (solid).
	The physical masses of the $\rho$, $a_2$ and $\rho_3$ mesons
	are indicated by stars, while the boxes represent lattice
	masses.\protect\cite{ORBIT}
	The darker shaded region represents the statistical error in the
	extrapolation, while the lighter region indicates an estimate of
	the systematic error associated with the fitting 
	procedure.\protect\cite{DMT}}
\end{center}
\end{figure}

In Fig.~4 we show the predicted $\rho$ Regge trajectory at
$m_\pi = 0.785$~GeV, corresponding to the strange quark mass, compared
with the trajectory at the physical light quark mass.
A fit through the central values of $\beta$ and the $\rho$
mass\cite{ORBIT} at $m_\pi = 0.785$~GeV yields a slope which is larger
than that of the trajectory at the physical quark mass, consistent with
the expected trend towards the heavy quark limit.

Although lattice data for the masses of orbital excitations are scarce,
there have been some pioneering calculations of the $a_2$ and $\rho_3$
meson masses by the UKQCD Collaboration,\cite{ORBIT} indicated by the
filled boxes in Fig.~4 (we use the fact that the $a_2$ trajectory lies
on top of the $\rho$ trajectory).
Comparing with the predictions from the PDF analysis, the calculated
$\rho_3$ meson mass lies within the predicted band, albeit within large
errors, while the $a_2$ mass lies on the edge of the predicted range.
Needless to say, further exploration of the masses of excited mesons
within lattice QCD would be very helpful in testing these predictions.

%%%%%%%%%%%%%%%%%%%%%%%%%%%%%%%%%%%%%%%%%%%%%%%%%%%%%%%%%%%%%%%%%%%%%%%%%
\section{Helicity Distributions}

Lattice simulations of the spin-averaged quark distributions provide
one of the benchmark calculations of hadron structure in lattice QCD.
Having established confidence in the reliability of the lattice
calculations through the correct treatment of the chiral behavior of
the moments, we can now turn to other PDFs of the nucleon.
Of particular importance are the helicity distributions, $\Delta q(x)$,
which describe the distribution of the nucleon spin amongst its quark
constituents.

While a complete determination of the helicity distributions requires
calculation of the singlet distribution, $\Sigma_q \Delta q$, for which
there have only been exploratory lattice calculations,\cite{SIGMA,FUK} a
more basic challenge remains to understand the axial vector charge,
$g_A$, given by the $n=0$ moment of the isovector $u-d$ distribution.
The results of several lattice calculations\cite{QCDSF,MIT,FUK} of
$g_A$ are compiled in Fig.~5.
(Not included are results from recent simulations using domain wall
fermions,\cite{VOLUME} which appear to have strong finite volume
dependence -- see below.)
When extrapolated linearly in $m_q$ to the physical quark mass, the
results are $\sim 10$--15\% lower than the experimental value.
Simulations with dynamical fermions\cite{MIT} (SESAM) yield results
consistent with the earlier unquenched calculations\cite{QCDSF,MIT,FUK}.

On the other hand, the behavior of the moments
$\langle \Delta x^n \rangle_{u-d}$ is known in the chiral\cite{JI} and
heavy quark limits, and can be used to constrain the extrapolation form,
as in Eq.~(\ref{xtrap}):
\begin{equation}
\label{xtrap_gA}
\langle \Delta x^n \rangle_{u-d}\
=\ \tilde a_n
   \left( 1 + \tilde c_{\rm LNA} m_\pi^2\log\frac{m_\pi^2}{m_\pi^2+\mu^2}
   \right)\
+\ \tilde b_n \frac{m_\pi^2}{m_\pi^2+\lambda_n^2}\ ,
\end{equation}
where $\tilde c_{\rm LNA} = -(1 + 2 g_A^2)/(4 \pi f_\pi)^2$, and
$\tilde b_n = 5/3^{n+1} - \tilde a_n (1 - \mu^2 \tilde c_{\rm LNA})$.
Using the same values of $\mu$ and $\lambda_n$ as in Fig.~1, the
results of the chiral extrapolation using Eq.~(\ref{xtrap_gA}) are
shown in Fig.~5.

\begin{figure}[htbp] 
\begin{center}
\epsfig{figure=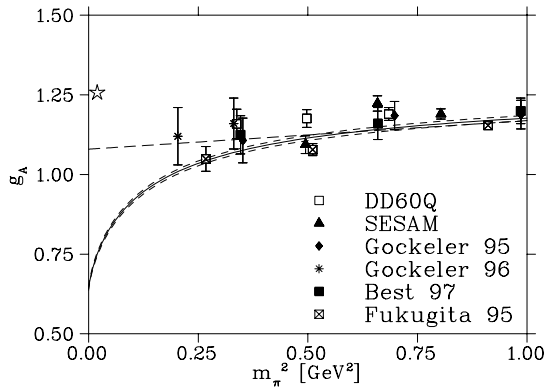,height=5.7cm}
\caption{Lattice data\protect\cite{QCDSF,MIT,FUK} on the nucleon axial
	charge, $g_A$, extrapolated linearly in $m_\pi^2$ (long dashed),
	and using Eq.~(\protect\ref{xtrap_gA}): best $\chi^2$ fit
	(solid), and fits to extrema of error bars (short dashed).}
\end{center}
\end{figure}

The result for $g_A$ is a downturn with decreasing $m_q$, making the
discrepancy with the empirical value larger.
The reasons for this could be severalfold.
Firstly, the chiral behavior in Eq.~(\ref{xtrap_gA}) arises only from
fluctuations $N \to \pi N \to N$, whereas it is well known that the
$\Delta$ plays an important role in $g_A$ through the fluctuation
$N \to \pi\Delta \to N$.
On the other hand, in the chiral limit the $\Delta$ has been
shown\cite{JI} to give vanishing chiral contributions at order
$m_\pi^2 \log m_\pi$, suggesting that higher order effects in the
chiral expansion are likely to be important.
The inclusion of these effects is currently under investigation.

More importantly, perhaps, to fully incorporate the pion cloud in a
lattice simulation a sufficiently large volume must be used.
Indeed, there are some indications\cite{VOLUME} of large finite volume
effects for $g_A$, which tend to {\em increase} $g_A$ in comparison
with the results from smaller volumes.
Clearly, in light of the results on the chiral extrapolation, it is
imperative to perform simulations on larger lattices to understand the
source of the discrepancy.

%%%%%%%%%%%%%%%%%%%%%%%%%%%%%%%%%%%%%%%%%%%%%%%%%%%%%%%%%%%%%%%%%%%%%%%%%
\section{Conclusion}

In this report we have highlighted the importance of model independent
constraints from the chiral and heavy quark limits of QCD in the
extrapolation of lattice data on parton distribution moments.
Inclusion of the nonanalytic structure associated with the infrared
behavior of Goldstone boson loops leads to a resolution of a
long-standing discrepancy between lattice data on low moments of the
spin-averaged $u-d$ distribution and experiment.

The importance of ensuring the correct chiral behavior is further
illustrated by comparing the $x$ distributions obtained by extrapolating
the lattice data using a linear and a chirally symmetric fit.
While the latter gives an $x$ distribution which is in quite good
agreement with the phenomenological fits, the linearly extrapolated data
give distributions with the wrong small-$x$ behavior,
which translates into a much more pronounced peak at $x \sim 1/3$,
reminiscent of a heavy, constituent quark--like distribution.
Our analysis suggests an intriguing connection between the small-$x$
behavior of the valence distributions and the $m_q$ dependence of meson
masses on Regge trajectories, which should be tested more thoroughly in
future simulations of the excited hadron spectrum.

Finally, we have highlighted the need for further study of moments of
the helicity distributions, and the axial vector charge of the nucleon
in particular, which appears to be underestimated in lattice simulations.
The inclusion of the pion cloud of the nucleon leads to a larger
discrepancy at the physical quark mass, indicating that lattice
artifacts such as finite volume effects may not yet be under control.
Further data on larger lattices, and at smaller quark masses, will be
necessary to resolve this issue.

%%%%%%%%%%%%%%%%%%%%%%%%%%%%%%%%%%%%%%%%%%%%%%%%%%%%%%%%%%%%%%%%%%%%%%%%%
\section*{Acknowledgements}

% W.M. would like to thank the organizers of the 3rd Circum-Pan-Pacific
% Symposium on ``High Energy Spin Physics''
% (Peking University, Beijing, Oct. 2001)
% for partial support.
We are grateful to D.B.~Leinweber, J.W.~Negele and D.B.~Renner for
helpful discussions.
This work was supported by the Australian Research Council,
and the U.S. Department of Energy contract \mbox{DE-AC05-84ER40150},
under which the Southeastern Universities Research Association (SURA)
operates the Thomas Jefferson National Accelerator Facility
(Jefferson Lab).

%%%%%%%%%%%%%%%%%%%%%%%%%%%%%%%%%%%%%%%%%%%%%%%%%%%%%%%%%%%%%%%%%%%%%%%%%

\end{document}